\RequirePackage{fix-cm}
\documentclass[twocolumn]{svjour3}          % twocolumn
\usepackage{graphicx}
\usepackage{amsmath}
\usepackage{amssymb}
\usepackage{xcolor}
\usepackage{epstopdf}
\usepackage{eqnarray}
\begin{document}

\title{{Continuous Vernier filtering of an optical frequency comb for broadband cavity-enhanced molecular spectroscopy}}



\author{Lucile Rutkowski         \and
        J\'er\^ome Morville %etc.
}


\institute{Lucile Rutkowski \at
              Department of Physics, Ume\aa University \\ 901 87 Ume\aa, Sweden\\
              \email{lucile.rutkowski@umu.se}           %  \\
           \and
           J\'er\^ome Morville \at
            Institut Lumière Matière, UMR5306 Université Lyon 1 - CNRS\\ Université de Lyon, 69622 Villeurbanne Cedex, France
}

\date{Received: date / Accepted: date}

\maketitle

\begin{abstract}
 We have recently introduced the Vernier-based Direct Frequency Comb Cavity-Enhanced Spectroscopy technique and we present the corresponding formalism for quantitative broadband spectroscopy. We achieve high sensitivity and broadband performance by acquiring spectra covering more than 2000 cm$^{-1}$ around 12600\,cm$^{-1}$ (800\,nm), resolving the 3$\nu$+$\delta$ band of water vapor and the entire A-band of oxygen in ambient air. 31 300 independent spectral elements are acquired at the second time scale with an absorption baseline noise of 2$\times$10$^{-8}$\,cm$^{-1}$ providing a merit figure of 1,1.10$^{-10}$\,cm$^{-1}$/$\sqrt{Hz}$ with a cavity finesse of 3000 and a cavity round-trip length around 3,3\,m. This state-of-the-art performance is reached through a continuous Vernier filtering of a Titanium:Saphire frequency comb with the cavity grid of resonances, obtained when the cavity free spectral range and the laser repetition rate are slightly mismatched. Here, we discuss the effect of the Vernier filtering on the measured absorbtion lineshape and we derive the formalism needed to fit molecular spectra.

\end{abstract}

\section{Introduction}
\label{intro}
Optical frequency combs (OFC), generated by femtosecond mode-locked lasers, are recognized to be very powerful devices for spectroscopy. They offer large bandwidth ($\Delta\nu$) discretized at the comb mode frequencies. They are defined by the repetition rate of the femtosecond oscillator $f_{rep}$ and by a carrier-envelop offset frequency $f_{ceo}$ translating the whole comb of a constant value comprised between $0$ and $f_{rep}$, each comb mode frequency being indexed by an integer $n$ and defined as $\nu_n^{las}=n. f_{rep}+f_{ceo}$. Several approaches \cite{Gohle2007,Nugent2012,Bernhardt2010,Grilli2012,Foltynowicz2013,Khodabakhsh2015,Zhu2014} have been designed to combine OFC's with the extended path lengths associated with high finesse cavities, to attain high sensitivity in molecular absorption spectra. Some resolve the comb mode structure (with 1 GHz mode locked laser) \cite{Gohle2007,Nugent2012}, some are fast (around or below the ms timescale of acquisition) \cite{Bernhardt2010,Grilli2012}, some particularly sensitive (baseline noise around 10$^{-4}$) \cite{Foltynowicz2013}, but none truly exploit the full bandwidth of OFC, restricting the accessible spectral range to roughly a few hundred of wavenumbers, around ten percent of the entire range of a typical Titanium:Sapphir (Ti:Sa) mode-locked laser.

Recently, we have developed a new scheme enabling the injection of the full spectrum of an OFC through a high finesse optical cavity and achieving high sensitivity, GHz resolution and sub-second acquisition time \cite{Rutkowski2014}. It uses Vernier filtering, whereby the laser repetition rate and the cavity free spectral range $FSR_{c}$ are deliberately mismatched (as are the two divisions of a Vernier caliper) so that the cavity outputs a new comb whose mode-spacing is sufficiently large that it can be dispersed with optical gratings. This Vernier coupling scheme is currently established as a calibration strategy for the astro-comb extensively developed to calibrate spectra recorded from telescope, creating the so-called astro-comb \cite{Wilken2012}. Gohle \emph{et. al.} \cite{Gohle2007} first applied it to laboratory molecular spectroscopy, resolving the comb mode structure of a 1 $GHz$ Ti:Sa mode-locked laser over 4 THz (130 cm$^{-1}$), in 10\,ms, but at the price of a poor sensitivity (baseline noise of a few 10$^{-6}$cm$^{-1}$ with an effective optical path-length around 300 m). In 2014 \cite{Zhu2014}, the same philosophy was extended to a 250 MHz Erbium doped fiber mode-locked laser \cite{Zhu2014} without the comb mode resolution (a resolution of 1,1 GHz is obtained). A spectral coverage of 160 cm$^{-1}$ and a sensitivity of 8$\times$10$^{-8}$\,cm$^{-1}$ (with an effective optical path-length of 11 km) was demonstrated on a time scale of 1 s. Still, only a small part of the OFC was exploited and sensitivity performances were again limited by the strong frequency-to-amplitude noise conversion resulting from the un-stabilized cavity with respect to the stabilized OFC.
Our approach is distinguished by the degree of mismatches between $FSR_{c}$ and $f_{rep}$. In astro-comb or previous work in molecular spectroscopy, the mismatch is large enough to filter out comb teeth adjacent to the tooth transmitted by the cavity. Each tooth of the new comb created at the cavity output thus corresponds to a single tooth of the original optical comb. To extract spectroscopic information from the sample placed inside the cavity, the cavity length is slightly swept to permit the transmission of the successive comb teeth through the cavity. The cavity transmission is thus modulated by the matching condition. Conversely, our scheme uses sufficiently small mismatches that several adjacent teeth are simultaneously (partially) transmitted. The matching condition is always fulfilled, and the cavity length sweep induces a continuous transmission. With a particular cavity locking scheme, this continuous Vernier filtering enables to probe the entire range of a free-running 100 MHz repetition rate Ti:Sa mode-locked laser, at GHz resolution and an optimized sensitivity. With this approach, an absorption baseline noise of 2$\times$10$^{-8}$\,cm$^{-1}$ is demonstrated with an effective optical path-length of 1,5 km and a spectral coverage larger than 2000 cm$^{-1}$ with 31300 independent spectral elements acquired in 1 s. We derive a full analytical model to describe the cavity transmission in presence of intracavity absorbing species for this continuous Vernier filtering regime. We show that both the real and imaginary part of the resonant molecular response need to be take into account revealing a strong dependency of the measured line profile on the sign of the Vernier mismatch. Using the full model and a non-linear fitting algorithm, we present results of the 3$\nu$+$\delta$ water vapor band adjustment over more than 100\,cm$^{-1}$ including more than 350 lines, and the measured spectra of the hot band of the magnetic dipole intercombination transition in molecular oxygen.

\section{The continuous Vernier filtering formalism}
\label{sec_continuousVernier}
\subsection{Identification of the Perfect Match}
\label{subsec_PM}

\begin{figure}[!t]

\includegraphics[scale=0.5]{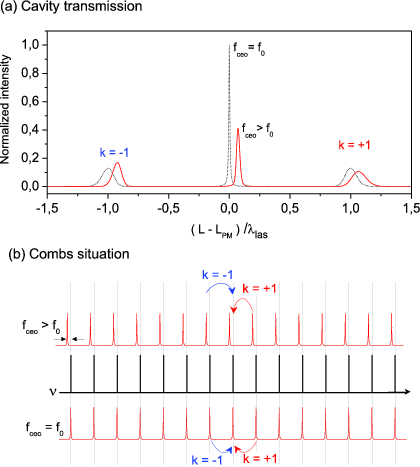}

\caption{Cavity transmission when the cavity length is swept over $\pm\lambda_{las}$ around $L_{PM}$ length at the PM, for matched offset frequencies $f_{ceo}=f_{0}$ (black dotted line) and mismatched (red line). (b) Frequency and resonance combs in both cases. Curved arrows indicate the path to reach the matching of the $(m=n\pm1)$th resonance with the nth tooth. In the case where $f_{ceo}>f_{0}$, $FSR_{c}$ is already reduced to reach the main peak and a lower reduction is required to match the $m+1$th resonance. Conversely, a larger increase of $FSR_{c}$ is required to match the $m-1$th resonance, reducing the number of teeth simultaneously transmitted and hence the optical power.}
\label{PM_and_2pks}
\end{figure}

The Vernier approach relies on a controlled mismatch between $FSR_{c}$ and $f_{rep}$ from the reference position corresponding to the perfect match (PM). In the frequency domain, this PM coupling occurs when both the scale and the origin of the OFC are tuned to match those of the cavity grid. In our system, $FSR_{c}$ (controlled by the cavity length), and $f_{ceo}$ are tuned to match $f_{rep}$ and the cavity offset frequency $f_{0}$ respectively. A large fraction of the OFC modes are transmitted simultaneously through the cavity, which can be expressed as $\nu_n^{las} = \nu_m^{cav}$ (with $n=m$) where the integer m is the order of the longitudinal cavity mode. This reference position is easily identified by applying a small cavity-length modulation, for instance with one of its mirrors mounted on a piezo-transducer (PZT). The global effect of this length-detuning is to dilate/compress the cavity grid, but at first order (if all intracavity dispersion sources such as mirror coatings or gas pressure are neglected) it shifts all the laser teeth out of resonance simultaneously. Filtering out the laser repetition rate, the optical power signal measured at the cavity output when a small dithering is applied around the PM resembles the signal obtained with a single frequency continuous-wave laser : a resonance curve corresponding to the Airy function (a Lorentzian line shape at high finesse) if the scanning speed guarantees the adiabatic response of the cavity, and the emergence of a more and more pronounced ringing when the scanning speed is increased \cite{Poirson1997}. For a cavity length (round-trip) detuning corresponding to $\pm\lambda_n^{las}$, the longitudinal cavity mode of order $m=n \pm 1$ reaches the laser mode $n$ and secondary peaks appears in transmission. Due to the slight $FSR_{c}$ variation, all the laser teeth can no longer enter in resonance simultaneously, and only a partial OFC is transmitted simultaneously through the cavity : if the equality $\nu_n^{las} = \nu_m^{cav}$ $(m=n\pm1)$ is satisfied for one laser mode, the following mode will be slightly shifted with its corresponding cavity resonance, and hence, partially transmitted through the Lorentzian line shape. The secondary transmission peak is thus weaker in intensity and broader than the PM position. This situation is schematically depicted on Fig.\ref{PM_and_2pks}. The pattern is symmetrical around the PM position. But if $f_{ceo}$ is not initially matched to the cavity offset, this symmetry is broken, resulting in a shifted and broadened main feature that is less intense than the PM peak and un-symmetrical neighboring peaks.

Once this PM position is identified, a cavity length variation $\Delta L$ of a significant number of laser wavelength which depends on the cavity finesse is applied. This creates a Moir\'e pattern between the laser comb and the grid of cavity resonances where some comb teeth are periodically coupled with a periodicity given by the mismatch $\Delta L$. The OFC cavity-filtered makes appear at the cavity output periodic and controllable Vernier coincidences defining the Vernier comb.

\subsection{The Vernier comb an its empty-cavity response function}
\label{subsec_VerniercombresponseFunc}

\begin{figure}[!]
\includegraphics[scale = 0.3]{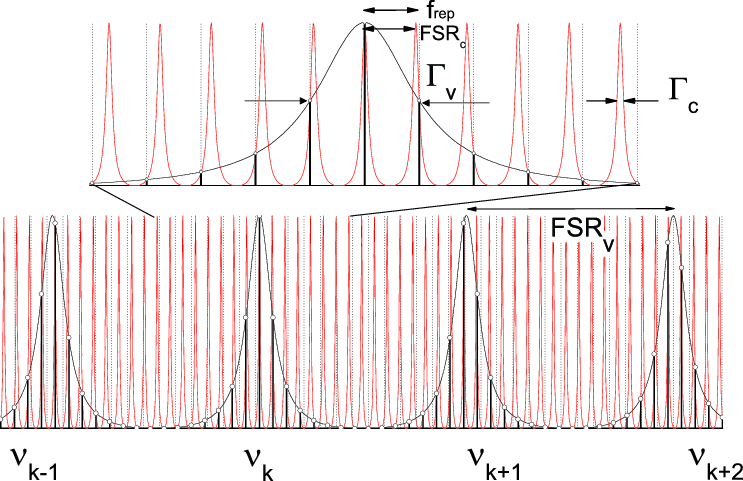}
\vspace{0.1cm}
\caption{Spectral envelope of the Vernier comb with a width $\Gamma_V$ determined by the cavity linewidth $\Gamma_c$ and the mismatch between $f_{rep}$ and $FSR_c$. Vernier orders are separated by $FSR_V$. For clarity, the cavity finesse is $F=7$ and the magnification factor $\mathcal{M}=15.2$.  }
\label{VernierComb}
\end{figure}

The Vernier comb and its response function can be derived with the cavity transfer function $H(\nu)$, expressed for the optical power as :
\begin{equation}\label{cavity transfer function}
H(\nu)  =  \cfrac{H_{max}}{1+(\frac{2F}{\pi})^{2} sin^2\left[\Phi(\nu)/2\right]}
\end{equation}
with the cavity finesse $F = \pi\cdot\sqrt{R}/(1-R)$ where $R$ is the round-trip amplitude reflection factor, and the maximum transmission $H_{max}  = \cfrac{T^{2}}{\left(1-R\right)^2}$, where $T$ is the product of the input and output coupler transmission amplitude. $\Phi(\nu)=2\pi\nu L_c/c - \phi_0$ is the phase accumulated over one round-trip, with $\phi_0$ a constant phase shift of the cavity inducing the cavity offset frequency $f_0=(\phi_0/2\pi)(c/L_c)$. We consider here neither absorption nor phase dispersion. At the perfect match all the comb frequencies are at perfect resonance, meaning $\Phi(\nu_n^{las})=n.2\pi$ for all integer n of the laser comb. This can be written as :
\begin{equation}\label{PM_roundtrip_phase}
2\pi.(n\cdot f_{rep}+f_{ceo})L_{PM}/c - \phi_0=n.2\pi
\end{equation}
from which round-trip length at the PM, $L_{PM}=c/f_{rep}$, and the laser frequency offset at the PM, $f_{ceo}=(\phi_0/2\pi).f_{rep}$ are deduced from the laser repetition rate. If the cavity round trip length is now varied by $\Delta L$, the round trip phase expressed at the comb frequencies takes the form :
\begin{equation}\label{roundtrip_phase_comb}
\Phi(\nu_n^{las})=2\pi.(n\cdot f_{rep}+f_{ceo}+\delta f_{ceo})(L_{PM}+\Delta L)/c - \phi_0
\end{equation}
where we deliberately introduced a possible variation, $\delta f_{ceo}$, of the laser frequency offset from its value at the PM. Finally, with the help of Eq.\ref{PM_roundtrip_phase}, the phase of Eq.\ref{roundtrip_phase_comb} (modulo $2\pi$) is rewritten as:
\begin{equation}\label{roundtrip_phase_Vernier}
\Phi(\nu_n^{las})/2\pi=\nu_n^{las}.\frac{\Delta L}{c}+\delta f_{ceo}.\frac{L_{PM}}{c}
\end{equation}
The cavity transfer function being described by the Airy function (Eq.\ref{cavity transfer function}), the comb tooth of order $n$ is coupled (at least partially) to a cavity mode when $\Phi(\nu_n^{las})<2\pi/F$. If furthermore $\Phi(\nu_{n+1}^{las})-\Phi(\nu_n^{las})-2\pi<2\pi/F$, adjacent teeth are also partially coupled to their respective resonance. In fact, using Eq.\ref{roundtrip_phase_Vernier}, this last relation can be expressed as $f_{rep}.\Delta L/c < 1/F$ or equivalently as $\Delta L<L_{PM}/F$. In this regime, injecting the round-trip phase expression of Eq.\ref{roundtrip_phase_Vernier} in the cavity transfer function (Eq.\ref{cavity transfer function}), a new Airy function is obtained. This function, plotted as a solid line in Fig.\ref{VernierComb}, has the following expression when sampled by the teeth of the OFC:
\begin{equation}\label{Vernier_Airy_function}
H_V(\nu)  =  \frac{H_{max}}{1+(\frac{2F}{\pi})^{2} sin^2\left[\pi(\nu.\frac{\Delta L}{c}+\delta f_{ceo}.\frac{L_{PM}}{c})\right]}
\end{equation}
and corresponds to the envelope of Vernier coincidences. It can be viewed as the response function of the Vernier filtering process, a comb of equally spaced Lorentzian profiles. Measuring the optical power transmitted through such a profile with a bandpass detector lower than the laser repetition rate eliminates the tooth structure and gives a weighted average frequency for the k'th order given by :\begin{equation}\label{order_frequency}
\nu_k^V = k\cdot \frac{c}{\Delta L} - \delta f_0\cdot\frac{L_{PM}}{\Delta L}
\end{equation}
corresponding to central frequency of the profile. The spacing between successive orders is the free spectral range of the Vernier comb and is given by $FSR_V=c/\Delta L$. The Vernier comb origin has an offset frequency, $f_{0V}$, directly linked to the offset frequency of the OFC with respect to the cavity. An important property appearing in view of Eq.\ref{Vernier_Airy_function} where the factor multiplying the sine function is unchanged with respect to Eq.\ref{cavity transfer function}, is that the cavity finesse also corresponds to the finesse of the Vernier comb. Hence, each profile has a linewidth $\Gamma_V$ given by $F=FSR_V/\Gamma_V$. For cavity finesses in the thousand range, a 100 MHz repetition rate mode-locked laser can provide Vernier order linewidth in the GHz range (ten successive teeth) with a Vernier comb easily spectrally resolved using standard optical diffraction devices as a moderate resolution of 1 THz is required here.

Eq.\ref{order_frequency} reveals an other important feature of the Vernier comb. Order frequencies of the Vernier comb are defined by just three free parameters : $f_{rep}$, setting the cavity length $L_{PM}$, the mismatch from this length $\Delta L$, and the difference between offset frequencies $\delta f_0$. Any drift or fluctuation of two of those free parameters can thus be balanced by the control of the third. This property will be exploited to lock the order frequency to a desired value.

Finally, introducing the factor $\mathcal{M}=f_{rep}/(FSR_c-f_{rep})=L_{c}/\Delta L\simeq L_{PM}/\Delta L$ as the inverse of the relative Vernier mismatch, all parameters of the Vernier comb, $\nu_k^V = k.FSR_V- f_{0V}$, and the profile linewidth $\Gamma_V$ can be written as :
\begin{align}
FSR_V &=\mathcal{M}.FSR_c\\
\Gamma_V &=\mathcal{M}.\Gamma_c \label{Gamma_V}\\
f_{0V} &=\mathcal{M}.\delta f_0
\end{align}
showing that, as long as $\mathcal{M}>F$, the Vernier coupling magnifies the cavity spectral structure probed by the OFC by the factor $\mathcal{M}$.

\subsection{Continuous Vernier filtering and its limit}
\label{subsec_continousVernierFilt}
 For an empty cavity, the complete expression of the power transmitted from the k'th Vernier order corresponds to the sum of all transmitted power by adjacent coupled teeth and is given by :
\begin{equation}\label{VernierPower_sum}
P(\nu_k^V)=\sum_{n=n_k-\mathcal{M}/2}^{n_k+\mathcal{M}/2} \frac{H_{max}}{1+\left(\frac{\nu_{n}^{las}-\nu_{k}^{V}}{\frac{\Gamma_V}{2}}\right)^{2} }\cdot S(\nu_{n}^{las})
\end{equation}
where the sine function of Eq.\ref{Vernier_Airy_function} has been linearized around the Vernier order frequency $\nu_k^V$ and the relation $F=FSR_V/\Gamma_v$ has been used. $n_k$ is the index of the laser tooth closest to $\nu_k^V$ and $S(\nu_{n}^{las})$ the spectral power of the laser comb. The sum is applied over half of the $FSR_V$ on each side of the Vernier order, which corresponds to $\pm \mathcal{M}/2$ in terms of teeth number.

If sufficient adjacent teeth are included under a Vernier order envelope, this sum expression can be written as an integral form as:
\begin{equation}\label{VernierPower_Int}
P(\nu_k^V) =  \int_{\nu_k^V-FSR_V/2}^{\nu_k^V+FSR_V/2}\frac{H_{max}}{1+\left(\frac{\nu-\nu_{k}^{V}}{\frac{\Gamma_V}{2}}\right)^{2} }\cdot S(\nu)\cdot\cfrac{d\nu}{f_{rep}}
\end{equation}
where the position of the laser teeth under the envelope does not appear anymore. The integral form thus provides a useful analytical expression to process the measured power as tooth position information is not required (the measured power is insensitive to the location of teeth within the envelope). Provided this integral form is appropriate, free-running mode-locked lasers where the whole comb is affected by spectral drift and noise can be used. We have already seen that this noise can be balanced with the cavity length to not affect the Vernier frequency. Now it also appears that it does not affect the transmitted power if the integral form applies. This is the main feature of continuous Vernier filtering. The power transmitted from an order when its frequency is scanned is only affected by the laser spectrum envelope variation; the OFC filtering process is continuous. Assuming a constant power per tooth $S_0$, the cavity transmission is constant and simply given by area of the Lorentzian profile : \\$\pi/2.H_{max}.S_0.\Gamma_V/f_{rep}=\pi/2.H_{max}.S_0.\mathcal{M}/F$, where the ratio $\mathcal{M}/F$ derived with the help of Eq.\ref{Gamma_V}, appears as the number of teeth included in the Vernier order linewidth.
\begin{figure}[!t]
\hspace{-0cm}

\includegraphics[scale = 0.5]{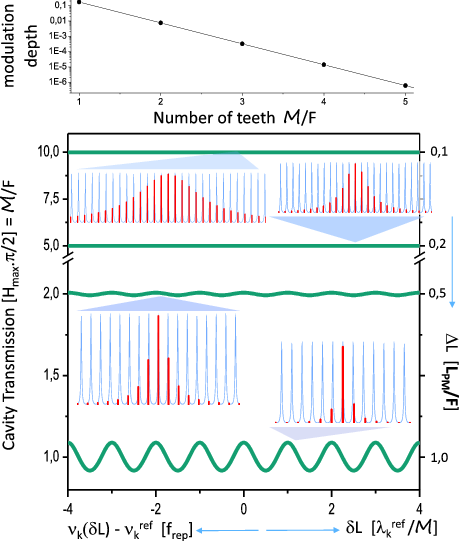}
\caption{Cavity transmission of the Vernier order k as a function of the number of teeth $\mathcal{M}/F$ under its profile linewidth when the frequency is scanned over $\pm 4.f_{rep}$ around the reference value $\nu_k^{ref}$. As fewer and fewer teeth are comprised in the order profile linewidth, a modulation appears in the transmission marking the $f_{rep}$ periodicity. Top panel reports the residual modulation amplitude as a function of $\mathcal{M}/F$. All quantity placed into brackets are the unit of the reported quantity. See the text for details and discussion.}
\label{Modulation}
\end{figure}

Obviously, in cavity enhanced molecular spectroscopy, the cavity transmission is not constant anymore and we will discuss in the section \ref{subsec ContVernFiltandabsorption} the modification of Eq.\ref{Vernier_Airy_function} in presence of intracavity absorption. In any case, it is useful to evaluate the limit of the continuous filtering process as it gives a range of validity and the highest resolution (the lower value of $\Gamma_V$) accessible with this approach. This has been done numerically and the result is plotted in Fig.\ref{Modulation}. For a chosen number of teeth $\mathcal{M}/F$ in the k'th order linewidth, the cavity transmission is simulated using Eq.\ref{VernierPower_sum} when the order frequency is scanned over few $f_{rep}$ around the reference frequency $\nu_k^{ref}$. The number $\mathcal{M}/F$ and the cavity finesse determine the mismatch length as $\Delta L= (\mathcal{M}/F)^{-1}.(L_{PM}/F)$ and the frequency scan is induced by a slight variation $\delta L$ around the round trip length $L_{PM}+\Delta L$. Eq.\ref{order_frequency} provides the needed length variation to scan $f_{rep}$, $\delta L=\lambda_k^{ref}/\mathcal{M}$, which itself depends on the mismatch. The power oscillates with a period of $f_{rep}$ when the Vernier order envelope slides across the laser teeth. It is clearly visible in Fig.\ref{Modulation} when few teeth form the envelope and it disappears progressively when the number increases. The average cavity transmission is given by the area of the Lorentzian curve and it corresponds directly to ratio $\mathcal{M}/F$ when it is normalized by $H_{max}.S_0.\pi/2$. The right axis of the graph indicates the value of the length mismatch $\Delta L$ for each curve, and it corresponds to the inverse of $\mathcal{M}/F$ when expressed in unit of $L_{PM}/F$. Finally, the graph above the figure shows the residual modulation amplitude as a function of the number of teeth where it appears that the amplitude reduction is fast and at $\mathcal{M}/F=5$, the modulation is already below $10^{-6}$. In practice, it is the level of experimental noise which sets the limit but it is hardly below this value and we can reasonably affect a number of five teeth in the linewidth of the Vernier order profile as the limit of our approach. For a 100 MHz mode locked laser, an ultimate resolution of 500 MHz is attainable.

\subsection{Vernier order frequency scanning speed}
\label{subsec_Vernierscanspeed}

\begin{figure}
\center
\includegraphics[scale = 0.25]{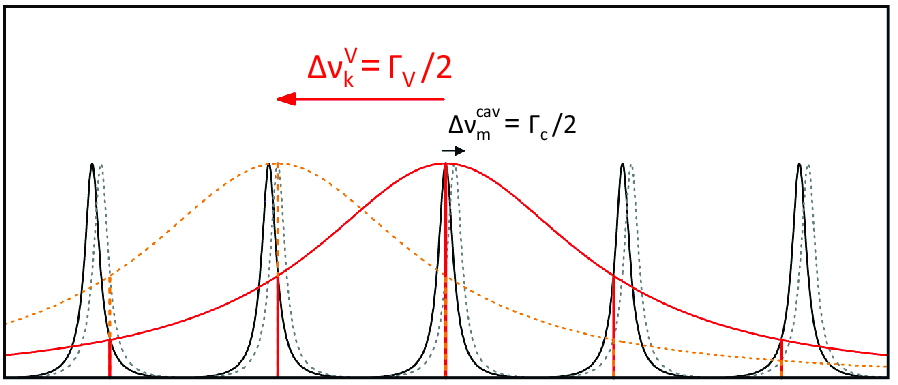}
\caption{Vernier order displacement (to the left) over $\Gamma_V$ when resonances are swept (to the right) over $\Gamma_c$.}
\label{fig_scanningspeed}
\end{figure}

Data acquisition rates are of course linked by the rate at which Vernier orders can be scanned. This limit is set by the condition needed to retrieve quantitative absorption measurement through the use of the cavity transfer function. In order to let the resonating optical field build up to its maximum value where the cavity transfer function applies, cavity resonances have to be swept across a laser tooth in a time longer than the cavity response time : $\Gamma_c^{-1}$. This defines the adiabatic cavity excitation and the adiabatic scanning speed, $W_{ad}=\Gamma_c^{2}$, when single frequency lasers are used. In the Vernier coupling case, the sweep of the resonance grid over a given spectral range induces a sweep of Vernier orders over a frequency range magnified by the magnification factor $\mathcal{M}$. As illustrated in Fig.\ref{fig_scanningspeed}, when resonances are swept over $\Gamma_c$, Vernier orders are swept over $\Gamma_V$. The adiabatic frequency scanning speed of Vernier orders thus corresponds to the adiabatic frequency scanning speed of the cavity magnified by $\mathcal{M}$ : $W_{ad}^{V}=\mathcal{M}.\Gamma_c^{2}$. This enables to cover extremely broad spectral range in a reasonable time.

\subsection{The Vernier comb response function with absorption}
\label{subsec ContVernFiltandabsorption}

\begin{figure}[!b]
 \vspace{-0.7cm}
\includegraphics[scale = 0.35]{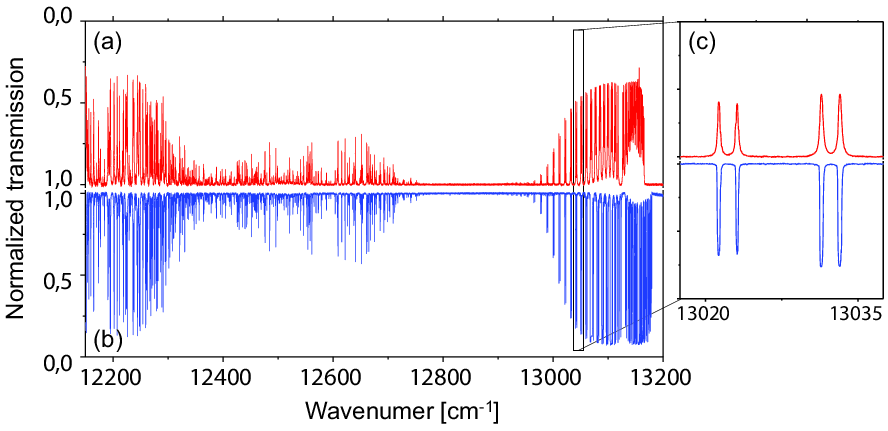}
\caption{Comparison between two measured spectra taken with Vernier mismatches of same absolute value ($|\Gamma_V|=2\, GHz$) but opposite sign: (a) positive case - inverted for clarity - and (b) negative case. A small spectral region is enlarged (c) to enlight the differences of lineshape and contrast between the two cases.}
\label{fig.signDL}
\end{figure}

When the cavity is filled with an analyte absorbing in the spectral range covered by the laser, the Vernier comb response function, which includes several cavity resonances, has to be modified in such a way that all effects affecting resonances are taken into account. Their amplitudes are reduced by absorption, their linewidths increased, but also their frequencies are shifted due to the small phase dispersion induced by resonant absorption. This phase dispersion stretches the cavity FSR under the line core and slightly compresses it into wings \cite{Chap5}. The main absorption effect on a Vernier order power is of course its reduction. However, in the continuous Vernier filtering limit, if effects are independently compared, the one on the amplitude reduces the power whereas the one on linewidth acts in opposite way as it slightly increases the teeth coupling and hence the power. The impact of $FSR_c$ variations is more subtle and appears to discriminate the sign of the Vernier mismatch. For instance, for negative values of $\Delta L$, $FSR_c$ is larger than $f_{rep}$. The FSR stretch into the line core thus further reduces the teeth coupling, and reduces the transmitted power. The situation is reversed for positive values where $FSR_c$ is lower than $f_{rep}$ resulting into a tooth-coupling increase into the line core. For both signs, the opposite discussion holds in the line wings. Experiment confirms that raw measured absorption profiles are dependent on the mismatch sign. Measured absorption profiles in the continuous Vernier filtering limit are not a straightforward convolution of the absorption line with the empty cavity response function. This effect is depicted in Fig. \ref{fig.signDL} showing some transitions recorded in ambient air taken with mismatch on either side of the PM. It is here clearly visible that contrasts and lineshapes are affected by the sign of the Vernier mismatch.

To derive the full Vernier response function, it is preferable to start here with the cavity transfer function expressed for the electric field :
\begin{equation}\label{field_transfer_function}
h(\nu)=\frac{T}{1-R.e^{-\frac{1}{2}\alpha(\nu).L_c}.e^{-i\Phi(\nu)}}
\end{equation}
where the absorption per unit length $\alpha(\nu)$ is introduced and where the round-trip phase now includes an absorption-induced dispersion term :
\begin{equation}\label{roundtrip_phase_and_abs}
\Phi(\nu)=\frac{2\pi\nu}{c}.(L_{PM}+\Delta L)-\phi_0-\phi_{abs}(\nu)
\end{equation}
The absorption-induced phase : $\phi_{abs}(\nu)=(2\pi\nu/c)\delta n(\nu)L_c$, is derived from the optical index variation $\delta n(\nu)$, induced by absorption which is itself deduced from the absorption coefficient through the Kramers-Kronig relations.
Eq. \ref{field_transfer_function} being evaluated at the comb frequencies around the k'th Vernier order, it can be rewritten in terms of the Vernier order frequency $\nu_k^V$ as :
\begin{equation}\label{field_transfer_function_bis}
h(\nu_n^{las})=\frac{T}{1-R.e^{-\frac{1}{2}\alpha(\nu_n^{las}).L_c-i[2\pi(\nu_n^{las}-\nu_k^V).\frac{\Delta L}{c}-\phi_{abs}(\nu_n^{las})]}}
\end{equation}
The argument of the exponential function is always small around $\nu_k^V$ and the denominator can be expanded to the first order as :
\begin{equation}\label{denominator1}
\begin{split}
(1-R)
&\left(1+
\frac{F}{2\pi}\alpha(\nu_n^{las})L_c\right)\\
&\left.
{} \left(1+i\cdot\frac{2\pi(\nu_n^{las}-\nu_k^V)\frac{\Delta L}{c}-\phi_{abs}(\nu_n^{las})}{\frac{\pi}{F}+\frac{1}{2}\alpha(\nu_n^{las})L_c}\right)\right.
\end{split}
\end{equation}

The last term of Eq \ref{denominator1} may be written
 \begin{equation}\label{denominator2}
1+i.\frac{\nu_n^{las}-\nu_k^V-\delta\nu_c(\nu_n^{las}).(L_c/\Delta L)}{\frac{1}{2}\left[\Gamma_c.(L_c/\Delta L)+\delta \Gamma_c(\nu_n^{las}).(L_c/\Delta L)\right]}
\end{equation}
where $\delta \Gamma_c=\alpha.c/2\pi$ corresponds to the increase of resonance linewidth induced by absorption, and $\delta\nu_c=(\phi_{abs}/2\pi).(c/L_c)=\nu.\delta n$ corresponds to the shift of the resonance frequency due to absorption.
The cavity transfer function for the optical power expressed at the comb frequencies in the vicinity of the k'th Vernier order can now be expressed as :

\begin{equation}\label{Power_Trans_func}
\begin{split}
H(\nu_n^{las})= \frac{H_{max}}{\left(1+\frac{F}{2\pi}.\alpha(\nu_n^{las}).L_c\right)^{2}}\cdot\\
&\hspace{-1.5cm}\left.{}\frac{1}{1+\left[\frac{\nu_n^{las}-\nu_k^V-\delta\nu_c(\nu_n^{las}).(L_c/\Delta L)}{\frac{1}{2}\left[(\Gamma_c+\delta \Gamma_c(\nu_n^{las})).(L_c/\Delta L)\right]}\right]^{2}}\right.
\end{split}
\end{equation}
and corresponds to the complete response function of the k'th Vernier order. The three effects of absorption are clearly identified, and this expression reveals that the two spectral features, linewidth and shift, of a cavity resonance, are here again magnified by the $\mathcal{M}=L_c/\Delta L$ factor. Moreover, it also shows that the sign of $\Delta L$ now affects the response function. Being quadratic, it does not impact the linewidth term, but the shift is sign-dependent. Using the factor $\mathcal{M}$, the Vernier response function can be written as :
\begin{equation}\label{Vernier_resp_func_abs}
H_V(\nu)= \frac{H_{max}}{\left(1+\frac{F}{2\pi}.\alpha(\nu).L_c\right)^{2}}\cdot \frac{1}{1+\left[\frac{\nu-\nu_k^V-\delta\nu_c(\nu).\mathcal{M}}{\frac{1}{2}\left[\Gamma_V+\delta \Gamma_c(\nu).|\mathcal{M}|\right]}\right]^{2}},
\end{equation}
where $\Gamma_V$ is now defined with the absolute value of $\mathcal{M}$ and, in the continuous Vernier filtering limit, the power transmitted from the k'th Vernier order as :
\begin{equation}\label{VernierPower_Int_abs}
P(\nu_k^V)=\int_{\nu_k^V-FSR_V/2}^{\nu_k^V+FSR_V/2} H_V(\nu)\cdot S(\nu)\cdot\cfrac{d\nu}{f_{rep}}
\end{equation}
Of course, expression \ref{VernierPower_Int} is retrieved if $\alpha(\nu)=0$.

Eqs.\ref{Vernier_resp_func_abs} and \ref{VernierPower_Int_abs} define the analytical function used to retrieve absorption spectrum from the measured cavity transmission. Once an absorption profile adapted to the thermodynamic conditions is selected, this function requires to apply the Kramers-Kronig relations to express $\delta\nu_c(\nu)$ as a function of the fitted parameters. As will be shown in the following, despite the fact that absorption varies significantly over the integration range, standard non-linear fitting algorithms efficiently converge to retrieve absorption spectra.
For extremely low absorption coefficient, however, expressions can be simplified by neglecting the tiny effect of absorption phase and by applying first order expansions to $\Gamma_V+\delta \Gamma_c(\nu).|\mathcal{M}|$ and the Vernier response function to be expanded as :
\begin{equation}\label{Vernier_resp_func_abs_lin}
\begin{split}
H_V(\nu)=H_{max}.&\left(1-\frac{F}{\pi}\alpha(\nu)L_c\right)\cdot\\
&\left.{}\frac{\Gamma_V+\delta \Gamma_c(\nu).|\mathcal{M}|}{\Gamma_V}\cdot
\frac{1}{1+\left(\frac{\nu-\nu_{k}^{V}}{\frac{\Gamma_V}{2}}\right)^{2} }\right.
\end{split}
\end{equation}
which can be further reduced to :
\begin{equation}\label{}
H_V(\nu)=\frac{H_{max}\left(1-\frac{F}{2\pi}\alpha(\nu)L_c\right)}{1+\left(\frac{\nu-\nu_{k}^{V}}{\frac{\Gamma_V}{2}}\right)^{2} }
\end{equation}
In this regime of extremely weak absorption, the fractional power reduction is then simply given by :
\begin{equation}\label{convolutionExp}
\frac{\Delta P(\nu_{k}^{V})}{P(\nu_{k}^{V})}=(\frac{F}{2\pi}\alpha L_c \ast g_V)(\nu_{k}^{V})
\end{equation}
corresponding to the intuitive cavity-enhanced absorption loss convoluted with the normalized Lorentzian profile of the Vernier order. The factor 2 results from the integration over the profile, similarly to what is obtained in integrated cavity enhanced approach \cite{Chap1}.

From a more general point of view, we see that the Vernier coupling in the continuous limit induces a situation with respect to resolved absorption lines which is almost equivalent to what is obtained when dealing with cavity resonance linewidth in the same range as the absorption linewidth. In this particular case which could be encountered in cavity-enhanced saturated-absorption spectroscopy, the absorption phase also distorts the spectral lineshape of the resonance. However, the sign of this phase is here controllable with the sign of the Vernier mismatch.

\section{Experimental setup}
\label{setup}
\begin{figure}[!b]
 \vspace{-0cm}
\includegraphics[scale = 0.35]{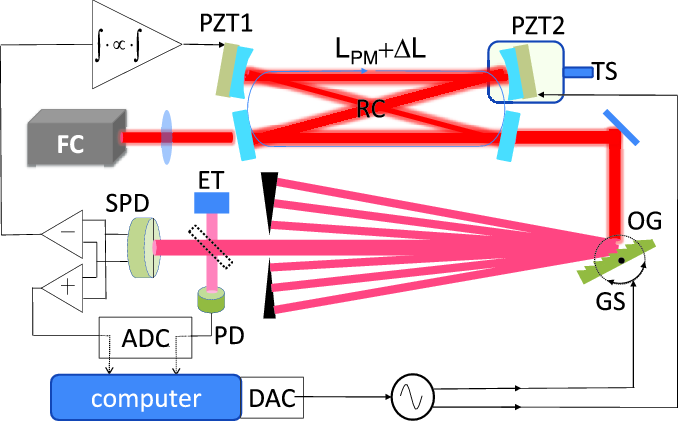}
\caption{Experimental setup. The Ti:Sa frequency comb (FC); RC, the bow tie ring cavity with the fast PZT-mirror (PZT1) used for the Vernier order frequency stabilization and the scanning mirrors PZT2 on which is applied the feedforward sinewave signal generated by the digital-to-analog-converter (DAC);  The optical grating (OG) supported by a galvo-scanner (GS) rotates with the same sinewave signal adapted in phase and amplitude; The Split-PhotoDiode (SPD) difference signal provides the Vernier order frequency error signal sent to the Integrator-Proportional-Integrator ; a low-finesse solid-\'etalon (ET) used with the photodiode (PD) provides the relative frequency scaling. This signal together with the summation SPD signal are acquired with a 16-bit analog-to-digital converter (ADC).}
\label{fig.setup}
\end{figure}

The set-up is described in Fig.\ref{fig.setup}. The FC is provided by a 90\,MHz mode-locked Ti:sapphire laser with a FWHM bandwidth of 30\,THz centered at $\lambda$=785\,nm delivering an average output power of around 0.5\,W. The beam is mode-matched to the open-air optical cavity formed by four broad-band dielectric mirrors in a bow tie arrangement, lowering high order modes excitation below the percent level. Input and output couplers have a nominal transmission of 0.1\,\% and dominate cavity losses, resulting in a cavity finesse $F=1000\,\pi$. One of the two other mirrors is mounted on a fast piezo-transducer (PZT1) with the holder described in \cite{Chadi2013}. The second is also mounted on a piezo-transducer (PZT2) with a travel range of 20\,$\mu$m and is supported by a translation stage. The cavity round-trip length is around 333\,cm, and the perfect-match length $L_{PM}$ is found by modulating the PZT2 with an amplitude of a few $\mu$m and by translating the stage, looking at a photo-detector signal temporarily placed just after the output mirror. From this position, the cavity length is adjusted to the desired value of $L_{PM}+\Delta L$ with the translation stage. The output beam is then sent onto a 1200\,l/mm optical grating which diffracts Vernier orders in the horizontal plane. This way, any variation of their frequency induces variation of their location in this plane. One of them is selected by a split-photodiode (SPD). The current difference of the two elements is amplified to provide a position error signal with a 100\,kHz bandpass. It results in a frequency error signal which is converted into a correction signal through a integrator-proportional-integrator and sent back to the fast PZT1. This closed-loop stabilizes the selected Vernier orders optical frequency at a 100\,MHz level with a bandpass of 20\,kHz filtering out efficiently thermal, mechanical and acoustic noise. A periodic frequency scan of the Vernier order is then obtained by applying an oscillating voltage to the galvo-scanner rotating the grating. The same voltage (adapted in phase and amplitude) is also applied to PZT2 as a feed-forward signal in order to reduce the range of the correction signal, which then essentially images relative fluctuations between the comb and the cavity. It can be noticed that a $\pi$-phase shift of the feed-forward and correction signals is applied when the Vernier mismatch sign is changed. The power transmitted from the Vernier order is obtained by the sum of the amplified currents of the SPD elements. This signal, translating the laser spectrum affected by the cavity enhanced absorption, is then sent to a 16-bit analog-to-digital converter (ADC) during a half period of the oscillating signal also used to trig the acquisition. The data treatment is processed in a Labview environment during the second half the period. In view of Eq.\ref{order_frequency}, the Vernier order frequency could, in principle, be deduced from the knowledge of $\Delta L$ during the scan. It also requires the precise knowledge of the offset frequency mismatch $\delta f_0$, of the laser repetition rate from which is deduced $L_{PM}$ and an accurate calibration of the PZT response. However, the cavity GDD, which depends on the used mirrors but also on the gas pressure, slowly varies the cavity free spectral range and must also be accurately determined to correct for the slight deviation it induces to Eq.\ref{order_frequency}. We circumvent those difficulties by sending a part of the selected beam onto a solid glass \'etalon without coating and a thickness of 1\,cm. The reflected beam measured with a photodiode is also sent to the ADC. Once normalized by the sum SPD signal, this \'etalon signal gives a quasi-sine function from which is extracted, during the data treatment time, a quasi-linear relative frequency scale with a wavenumber sampling around 0,333\,cm$^{-1}$. This way, the cavity transmission on a relative frequency scale is obtained at each period of the oscillating voltage applied to the galvo-scanner and can be conveniently time-averaged. The absolute frequency scale of an averaged spectrum is deduced by selecting two observed lines referenced in the HITRAN database \cite{HITRAN2012} and placed on each side of the spectrum. It can be noticed that due to the broad spectral coverage, a linear absolute frequency scale can only be retrieved if the spectral dependance of the \'etalon optical index is taken into account.

\begin{figure}[h!]
\hspace{-0cm}
\includegraphics[scale=0.35]{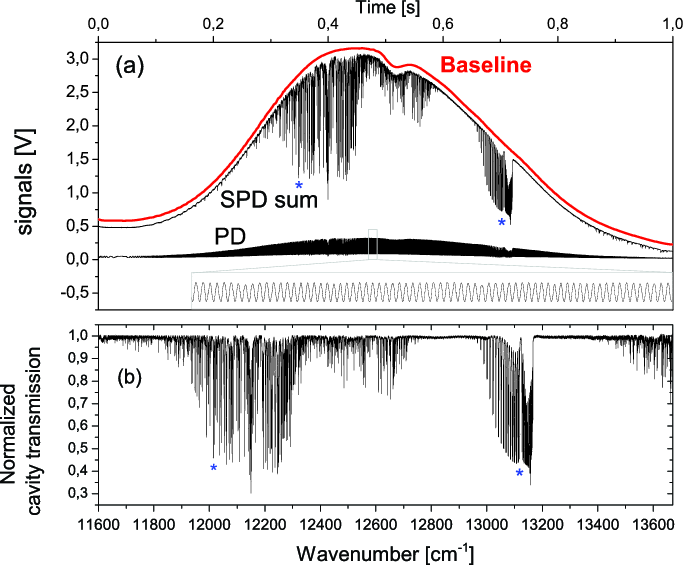}
\caption{(a) Raw signals provided by the SPD summation and the \'etalon PD when a Vernier order is scanned over the full bandwidth of the laser with a length mismatch $\Delta L$ of +50\,$\mu$m. The calculated baseline retrieved from the raw spectrum is also plotted with a vertical offset for clarity. The inset shows the relative frequency ruler provided by the \'etalon reflection. The absolute frequency scale is determined using the two transitions identified with the HITRAN database and marked with stars. (b) The normalized cavity transmission resulting from raw data and the frequency calibration procedure spanning more than 2000\,cm$^{-1}$. The final spectrum of ambient air covers the entire weak $3\nu+\delta$ band of water vapor ranging from 11600 to 12800\,cm$^{-1}$ and the beginning of the $4\nu$ band starting at 13400\,cm$^{-1}$. Between those two water bands, the full oxygen A-band is also measured.}
\label{rawdata}
\end{figure}

In Fig.\ref{rawdata} (a) are depicted the measured signals from the SPD and the \'etalon PD acquired during a half period of the Vernier sweep when the length mismatch is set to $\Delta L=+50\,\mu m$. The SPD signal describes the laser spectrum affected by the cavity-enhanced absorption of ambient air while the \'etalon PD provides a quasi-sinewave signal (enlarged in the inset) used as a relative frequency ruler. The laser spectrum without absorption is determined with a baseline fitting algorithm applied to the SPD signal. Those steps enable to retrieve the normalized cavity transmission spectrum over the full bandwidth of the FC. The final spectrum of ambient air covers the entire weak $3\nu+\delta$ band of water vapor ranging from 11600 to 12800\,cm$^{-1}$ and the beginning of the $4\nu$ band starting at 13400\,cm$^{-1}$. Between those two water bands, the full oxygen A-band corresponding to the $X^{3}\Sigma^-_g \rightarrow b^{1}\Sigma^+_g $ magnetic dipole intercombination transition is also identified. The frequency $f_{scan}$ of the sinewave applied to the galvo scanner is set by the probed optical range and the adiabatic scanning speed $W_{ad}^{V}$ which depends on the cavity finesse. For a finesse value around 3000, a cavity length of 333\,cm, a mismatch of 50\,$\mu$m and a spectral coverage of 2000\,cm$^{-1}$, $f_{scan}$ should not exceed 0.5\,Hz which set the acquisition time to 1\,s. Also, the spectral resolution of the Vernier order ($c/(\Delta L\!\cdot\!F)$) is 2\,GHz (corresponding to 0.067\,cm$^{-1}$) with those parameters providing 31300 independent spectral elements in 1\,s. Of course, for such a broadband spectrum, the finesse cannot be assumed to be constant over the whole range. However, based on a finesse measurement outside the molecular bands using the ringing method \cite{Poirson1997} in well agreement with the specifications provided by the constructor, the finesse is expected to remain within a variation of less than 20\,\%.

\begin{figure}[h!]
\hspace{-0.0cm}
\includegraphics[scale=0.4]{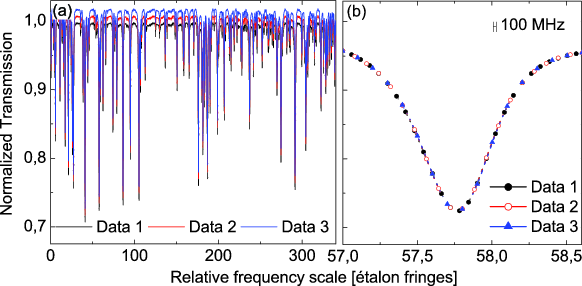}
\caption{(a) Three successive normalized transmissions measured whithout unlocking the system and each of them calibrated regarding the fringes produced by a 1\,cm thick \'etalon (vertical offset for clarity). (b) Enlargement of one line, illustrating the good reproducibility of the measurement and the frequency scale precision at the level of 100\,MHz (marker).}
\label{Reproducibility}
\end{figure}
After acquisition, the cavity transmission calibrated against the \'etalon ruler can be averaged thanks to the good stability provided by the correction loop.
To demonstrate the precision of the relative frequency axis over several measurements, three broadband normalized cavity transmissions are shown in Fig.\ref{Reproducibility}(a) (offseted for clarity), where it is already visible that no major frequency shift is observed. The three spectra are superimposed and enlarged in panel (b) together with a benchmark of 100\,MHz, confirming the previous statement of a spectral stabilization at 100\,MHz level. Thus the successive spectra can be efficiently averaged several times without being affected by the drift of the \'etalon fringes due to the low frequency thermal noise.

\section{Results and performances}

\begin{figure}[t!]
\hspace{-0.0cm}
\includegraphics[scale=0.42]{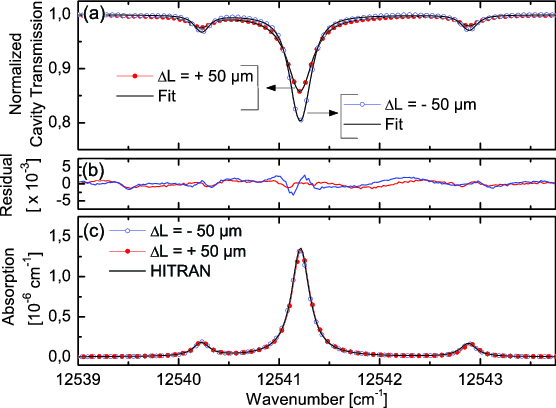}
\caption{(a) Normalized spectra (100 average) covering 3 water transitions measured at opposite length mismatches (red markers $\Delta L$ = +\,50\,$\mu$m, blue markers $\Delta L$ = -\,50\,$\mu$m) superimposed on their respective fits calculated with a cavity finesse of 3150 (black solid curves). (b) Residuals of the positve mismatch fit (in red) and of the negative mismatch fit (in blue). (c) Comparison between the HITRAN absorption spectrum calculated for a relative humidity (RH) of 45\,$\%$ (black solid curve) and the absorption spectra retrieved from the fits of the transitions frequencies, linestengths and total linewidths (negative mismatch with blue markers, positive mismatch with red markers). The agreement between the different lines parameters is better than 2\,$\%$, at the level of precision of the fits' residuals.}
\label{posneg}
\end{figure}

Molecular spectroscopic information, such as width, position and strength of atmospheric transitions, can be deduced from the normalized cavity transmission using the formalism developed in section \ref{subsec ContVernFiltandabsorption}. Especially, Eqs. \ref{Vernier_resp_func_abs} and \ref{VernierPower_Int_abs} show that the measured signal is not a simple convolution of the absorption line with the Vernier order profile, and particularly that the sign of the length mismatch affects the result. To check the model validity, two averaged cavity transmissions were recorded with opposite length mismatches as already shown in Fig.\ref{fig.signDL} and an isolated line, referenced in the HITRAN database \cite{HITRAN2012}, was selected out of the weaker part of the water band around 12541\,cm$^{-1}$. During the baseline correction, two additional sinewaves were added to the baseline algorithm to correct for the small \'etalon fringes induced by two 3\,mm thick cavity mirrors. Both normalized spectra are superimposed on Fig.\ref{posneg}(a). As previously mentioned, the positive mismatch exhibits a reduced contrast compared to the negative one. Small lines are also visible on both sides. Both spectra were adjusted using a non-linear algorithm developed in a Matlab environment based on Eq.\ref{VernierPower_Int_abs} which is replaced by :

\begin{equation}\label{Fit_function}
P(\nu_k^V)=\int_{\nu_k^V-FSR_V/2}^{\nu_k^V+FSR_V/2} \frac{H_V(\nu)}{H_{max}}\cdot\cfrac{d\nu}{f_{rep}}
\end{equation}
as the slowly varying laser spectrum can be taken out of the sum and disappears in the experimental normalization process. As the probed gas is at atmospheric pressure, a purely collisional regime is assumed to govern the absorption profile which is then by a Lorentzian profile.
%FWHMcollisionel=0.21 wn et FWHMDoppler=0.037wn
In this case, the shift of the resonance frequency due to absorption $\delta\nu_c(\nu)$, involved in Eq.\ref{Vernier_resp_func_abs}, takes the following expression :
\begin{equation}\label{KK_Lorentz}
\delta\nu_c(\nu)=\sum_{a=1,2,3}{\frac{c}{2\pi}\cdot\frac{\nu_a-\nu}{\Gamma_a}\cdot\alpha_a(\nu)}
\end{equation}

Both measurements are fitted and the adjusted curves are superimposed on the measured spectra in Fig.\ref{posneg}(a). For both mismatch signs, measurement and fit are indistinguishable demonstrating the ability of the derived model to apprehend the whole Vernier behavior. The residuals of the fits are plotted in Fig.\ref{posneg}(b) and confirm the good agreement between data and model. Appart from some remaining \'etalons, the distortions exhibited by the residuals in the center of the lines are whithin the expected residual distortion induced by the simplistic Lorentzian profile that would be of the order of 3\,$\%$ of the absorption line contrast at atmospheric pressure, actually showing the need to switch for a more sofisticated line profile such as Voigt good up to 3\,10$^{-3}$ precision of the line at atmospheric pressure. The result of the fits gives access to the three lines' linestrengths, linewidths and transition frequencies. Fig.\ref{posneg}(c) shows the comparison of the absorption spectra recalculated from the fitted line parameters for both mismatch and the absorption spectrum calculated using the HITRAN database parameters and the relative humidity (RH) measured using a commercial hygrometer. Both spectra compare well with HITRAN prediction.

\begin{figure}[!b]
\hspace{-0.0cm}
\includegraphics[scale=0.43]{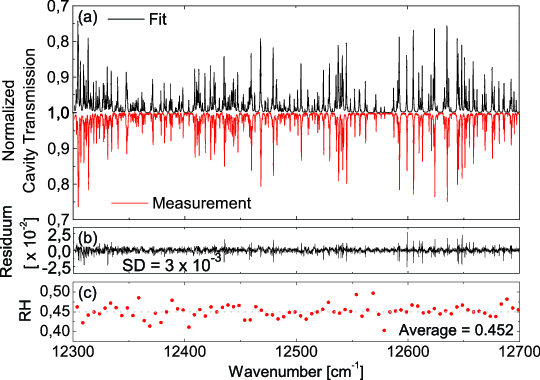}
\caption{(a) Normalized water spectrum (in red - no averaging) aquired with a resolution of 2\,GHz taken on the positive mismatch side together with a fit of the water concentration (in black) based on the HITRAN parameters and a lorentzian absorption profile, illustrating our ability to deal with congested spectra. (b) Residual of the fit. The residuum standard deviation (SD) calculated on the full spectral range is equal to 3$\times$10$^{-3}$, demonstrating the high sensitivity of the system. The average peak to peak value stays within the limit expected from the lorentzian profile approximation. (c) Relative humidity retrieved from the fit, indicating an average humidity of 45.2\,$\%$ that agrees well with the laboratory measurement of 45\,$\%$ using a commercial device (dashed grey line).}
\label{H2Obroadband}
\end{figure}

The same model is applied to a spectrum covering a broader range, shown in Fig.\ref{H2Obroadband}(a), extracted from the full spectrum already shown in Fig.\ref{rawdata}. It contains 358 water absorption lines that are fitted using the HITRAN parameters to retrieve the water concentration. As previously mentionned, the finesse spectral variations cannot be neglected on such a large spectral coverage. Taking it into account, the whole spectrum is fitted piece by piece with a step of 5\,cm$^{-1}$ to retrieve the accurate concentration distribution. The equally good quality of the fit is assessed with the residuum in panel (b). The sharp peaks present in the residual are attributed to the Lorentzian approximation and the 100\,MHz instability of the frequency axis. The resulting RH distribution is plotted in panel (c) - the saturation pressure at 23\,$^\circ$C is assumed to be equal to 0.0277\,atm. The average RH is equal to 45.2\,$\%$, which compares well with the RH value of 45\,$\%$ measured in the laboratory. Final consideration on this spectrum is about the overall sensitivity demonstrated here that can be express in terms of figure of merit (FM) defined as the noise equivalent absorption (NEA) normalized by the square root of the acquisition time of a single spectral element. The standard deviation of the residuum calculated over the 400\,cm$^{-1}$ is equal to 3$\cdot$10$^{-3}$. The SD can be converted linearly to a NEA via the usual cavity-enhancement relation for integrated measurement \cite{Chap1} $NEA = SD\times 2\pi/(F\!\cdot\! L)$, leading to a NEA of 2$\cdot$10$^{-8}$\,cm$^{-1}$. The 2100\,cm$^{-1}$ coverage of the whole spectrum (Fig.\ref{rawdata}) were aquired in 1\,s with a resolution of 0.067\,cm$^{-1}$, which makes the acquisition time of one spectral element equal to 32\,$\mu$s. The FM is then equal to 1.1$\cdot$10$^{-10}$\,cm$^{-1}/\sqrt{Hz}$.

\begin{figure}
\hspace{-0cm}
\includegraphics[scale=0.375]{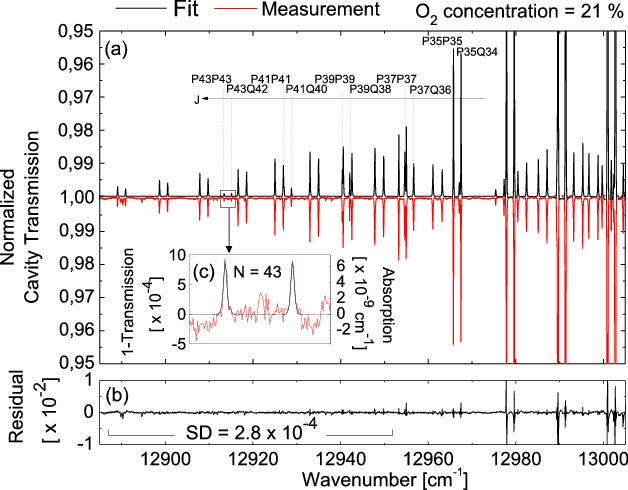}
\caption{(a) Normalized cavity transmission spectrum (in red - 100 averaging) acquired with a resolution of 2\,GHz taken on the positive mismatch side ($\Delta$L = +\,50\,$\mu$m) together with a fit of the $O_2$ concentration (in black) based on the HITRAN parameters and a lorentzian absorption profile. (b) Residual of the fit with a standard deviation of 3$\times$10$^{-4}$. The inset (c) shows a enlargement of the weakest O$_2$ doublet around 12014\,cm$^{-1}$. In this regime, the Vernier transmission is purely linear and the absorption scale is shown on the right axis of the inset, demonstrating our ability to measure absorption lines of 7$\cdot$10$^{-9}$\,cm$^{-1}$ with a signal to noise ratio close to 2.}
\label{O2sensitivity}
\end{figure}

To further assess the sensitivity of our system, a 100 time-averaged cavity transmission was acquired around 12945\,cm$^{-1}$, covering 120\,cm$^{-1}$ (Fig.\ref{O2sensitivity}). It was again measured with a resolution of 2\,GHz and for a positive sign of the mismatch. Again, two sinewaves were added to the baseline normalization procedure in order to correct for the mirror-induced \'etalon fringes. The normalized cavity transmission (in red) is plotted together with a fit (in black - panel (b)) and residuum (panel (c)). The noise level of 3$\cdot$10$^{-4}$, calculated over the residuum part that is not affected by the lorentzian profile distortion, confirms the white noise averaging effect since it is reduced by a factor 10 compared to the previous value obtained from a single acquisition spectrum (Fig.\ref{H2Obroadband}(c)). This spectral ranges contains the low energy tale of the $X^{3}\Sigma^-_g \rightarrow b^{1}\Sigma^+_g $ doubly forbidden band of $O_2$, interleaved with the hot band of the same system, three order lower in intensity. The oxygen concentration retrieved from the fit agrees within 0.5\,$\%$ with the expected concentration in atmosphere at sea level. To illustrate this high sensitivity, the inset Fig.\ref{O2sensitivity}(c) focuses on the weakest oxygen doublet measured in the fundamental band, that exhibits a signal to noise ratio close to 2, giving access to rotational quantum number up to $N=43$. It is plotted with both absorbance (on the left of the inset) and absorption (on the right of the inset) scales, converted using the same linear relation as the NEA. It demonstrates our ability to measure absorption line as weak as 7\,$\cdot$10$^{-9}$\,cm$^{-1}$. Eventually, this last result illustrates the ability of this approach to measure and quantitatively exploit absorption lines on a dynamic range of three orders of magnitude.

\section{Conclusions and outlook}

We have developed a formalism that comprehends the whole behavior of the Vernier coupling of a frequency comb into a mismatch cavity. This formalism enables to adjust spectra taken for both sign of the mismatch with the same good agreement, describing the observed discrepancy between both sign contrasts with a reliable accuracy. The cavity dispersion, that is one of the major bandwidth limitations of other direct frequency comb cavity-enhanced spectroscopy techniques, does not have to be considered in this formalism. This is enlighten by our ability to measure spectra that cover 2100\,cm$^{-1}$, which would be impossible with any other tight-locked technique. The Vernier order filtered by the cavity is yet tightly locked with a simple locking technique which is both robust and inexpensive. The combination of the grating with the split photodiode exploited here effectively stabilize the cavity resonances in regards of the comb modes at 100\,MHz level, already enough to be able to average the successive spectra. This gives high signal to noise ratio, which actually makes essential to implement the Voigt profile in our model to describe with better veracity the actually Doppler broadened measured lineshapes.

The overall performance gives a figure of merit equal to 1.1$\cdot$10$^{-10}$\,cm$^{-1}/\sqrt{Hz}$ which compares well with the other comb-based spectroscopy techniques, and absorption detection down to the level of 7$\cdot$10$^{-9}$\,cm$^{-1}$ has been demonstrated. The technique only require the additional measurement of the cavity finesse and of the length mismatch $\Delta$L to precisely calibrate the absorption scale. On the other hand, the frequency calibration is realized with the relative frequency ruler provided by the low finesse \'etalon, that contribute to the simplicity of the setup and is already enough precise for the presented measurements. Yet, this calibration could be made more accurate by replacing the solid \'etalon by a stabilized Michelson interferometer and by adding an absolute reference in the OFC beam path.

Further developments of the spectrometer include a vacuum cell containing the cavity in order to reliably control the probed gas settings, and to measure accurately the cavity finesse by ringdown measurement over the whole spectral range. This will also enable to measure the real background of the absorption spectrum, thus opening up to the detection of broad spectral features such as large molecules or aerosols. Anyhow, the Vernier spectrometer is already a reliable, simple and low-cost device delivering high sensitivity molecular spectra on a spectral range only limited by the laser bandwidth. We believe it can find potential application in exploratory spectroscopy, multispecies detection for atmospheric and climate research.

%Text with citations \cite{RefB} and \cite{RefJ}.
%\subsection{Subsection title}
%\label{sec:2}
%as required. Don't forget to give each section
%and subsection a unique label (see Sect.~\ref{sec:1}).
%\paragraph{Paragraph headings} Use paragraph headings as needed.
%\begin{equation}
%\end{e&quation}

%\begin{figure*}
%  \includegraphics[width=0.75\textwidth]{example.eps}
%\end{figure*}
%\begin{table}
%\begin{tabular}{lll}
%\hline\noalign{\smallskip}
%\end{tabular}
%\end{table}

%\begin{acknowledgements}
%\end{acknowledgements}



\end{document}